\documentstyle[prl,aps,epsfig,multicol]{revtex}

\begin{document}

\title{Competition of Spin-Fluctuations and Phonons in Superconductivity of ZrZn$_2$}
\author{D.J. Singh and I.I. Mazin}
\address{Center for Computational Materials Science,\\
Naval Research Laboratory, Washington, DC 20375, U.S.A.}
\date{\today}
\maketitle

\begin{abstract}
It has been long suspected that spin fluctuations 
in the weak itinerant ferromagnet ZrZn$_2$
may lead
to a triplet superconductivity in this material. Here we point out
another possibility, a spatially inhomogeneous singlet 
superconducting state (a Fulde-Ferrell-Larkin-Ovchinnikov state).
We report detailed electronic structure calculations, as well as 
calculations of the zone center phonons and their coupling with electrons.
We find that the exchange splitting is nonuniform and may allow
for gap formation at some parts of the Fermi surface. We also
find that there is substantial coupling of Zr rattling modes with
electrons, which can, in principle, provide the necessary pairing
in the s-channel.
\end{abstract}
\begin{multicols}{2}

The recent discovery of superconductivity in the ferromagnetic phase of ZrZn$_{2}$%
\cite{Peid} has revived interest in this unusual compound. ZrZn$_{2}$
has traditionally been considered a prototypical example of a weak itinerant
(Stoner) ferromagnet. Very small magnetic moments (values from 0.12 to 0.23 $%
\mu _{B}$) have been reported. These do not saturate even at magnetic
fields as high as 35 T, indicating softness of the magnetic moment
amplitude and suggesting existence of soft longitudinal spin fluctuations.
Already in the first report of ferromagnetism\cite{matt} it was pointed out
that ZrZn$_{2}$ may be a ferromagnetic superconductor, of the type discussed earlier by
Ginzburg \cite{vl}. Later the idea  of triplet spin-fluctuation induced
superconductivity in ferromagnetic ZrZn$_{2}$ was theoretically elaborated by Fay and
Appel\cite{fa}. However, experimental searches for 
superconductivity in ZrZn$_{2}$ \cite{search} were unsuccessful, till
now.
It is therefore very tempting to identify 
ZrZn$_{2}$ as a  triplet superconductor. This is supported by the fact
that superconductivity seems to disappear with pressure at about the same point
where the magnetic Curie temperature goes to zero. On the other hand, the low
superconducting fraction does not allow accurate determination of the critical pressure
for the superconducting transition to clearly show that it coincides with that
for the magnetism. Thus one cannot fully exclude the possibility that the two types of orders
are spatially separated. Still, it is likely that they do coexist, and if
so, a key question is whether their interaction is
constructive or destructive. A definite answer will require more
study. However, before making any theoretical speculations one needs to get a
detailed understanding of the electronic structure of this compound, and its
relation with superconductivity and magnetism. Here we present an accurate
analysis of the para- and ferromagnetic electronic structure and identify
what seems to be the most interesting and relevant 
features of the band structure for superconductivity. 

We find that, despite the very
small magnetization, the rigid band model is inappropriate for ZrZn$_{2}.$
The exchange splitting, $\Delta_{xc}({\bf k})$,
 differs from band to band by more than a factor of two.
Even more importantly, some bands have extremely small ($\Delta k\approx 0.12$ 
\AA $^{-1})$ exchange splitting in $k$-space, making superconductivity in these bands
less sensitive to magnetic ordering. Furthermore, the calculated Fermi
surface shows substantial nesting features, suggesting that antiferromagnetic
fluctuations
may play a role in superconductivity. Finally, we found that the rattling Zr modes
are soft and couple strongly with electrons. We emphasize that, while  triplet superconductivity is
an interesting possibility, the facts are nevertheless compatible with an
inhomogeneous 
Fulde-Ferrell-Larkin-Ovchinnikov (FFLO) state\cite{fflo} of the s-wave symmetry as well.

\begin{figure}[tbp]
\vspace{-.2in}
\centerline{\epsfig{file=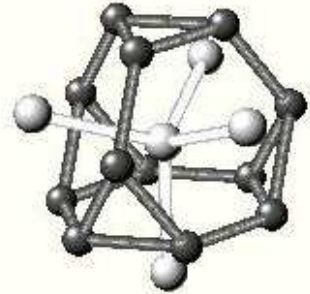,width=0.6\linewidth}}
\vspace{-.1in}
\setlength{\columnwidth}{3.2in} \nopagebreak
\caption{Coordination of a Zr atom in ZrZn$_2$.}
\label{str}
\end{figure}
ZrZn$_{2}$ occurs in a C15 crystal structure, where the Zr forms a diamond-type
lattice, and the Zn forms a network of corner-sharing tetrahedra (another way
to visualize this structure is to imagine a spinel without anions).  An
interesting aspect is the ``overcoordination'' of Zr: it has
16 nearest neighbors consisting of 12 Zn, forming a truncated tetrahedron (Fig.\ref{str}),
 at a distance (at T=50K\cite{str})
5.745 a.u., and 4 Zr at 6.001 a.u. The latter approximately
corresponds to the bond length in Zr metal. Since the metallic radius of Zn
is 16\% smaller than that of Zr, Zr and Zn do not form strong bonds. On the
other hand, Zr, unlike carbon, does not form highly directional bonds, so 4 Zr-Zr
bonds in ZrZn$_{2}$ do not provide strong bonding either. This makes Zr
``rattling'' rather soft. At the same time, Zn has 8 neighbors at a distance
4.9 a.u., noticeably {\it less} than in Zn metal (5.04 -- 5.51 a.u.).
Correspondingly, one expect that
Zn bond-stretching vibrations should be relatively hard.

\begin{figure}[t]
\vspace{-.2in}
\centerline{\epsfig{file=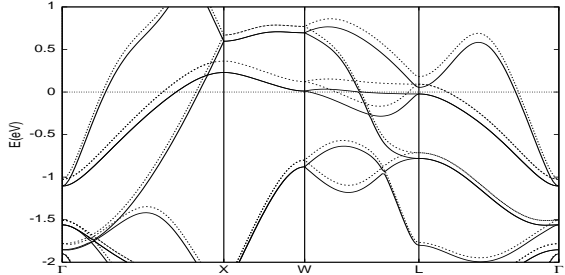,angle=270,width=0.88\linewidth}}
\vspace{-.5in}
\setlength{\columnwidth}{3.2in} \nopagebreak
\caption{LSDA band structure of ZrZn$_2$.  $E_F$ is at 0. Solid (dashed)
lines show majority (minority) bands corresponding to the
magnetization of 0.2 $\mu_B$/f.u.}
\label{bands}
\end{figure}
\begin{figure}[tbp]
\centerline{\epsfig{file=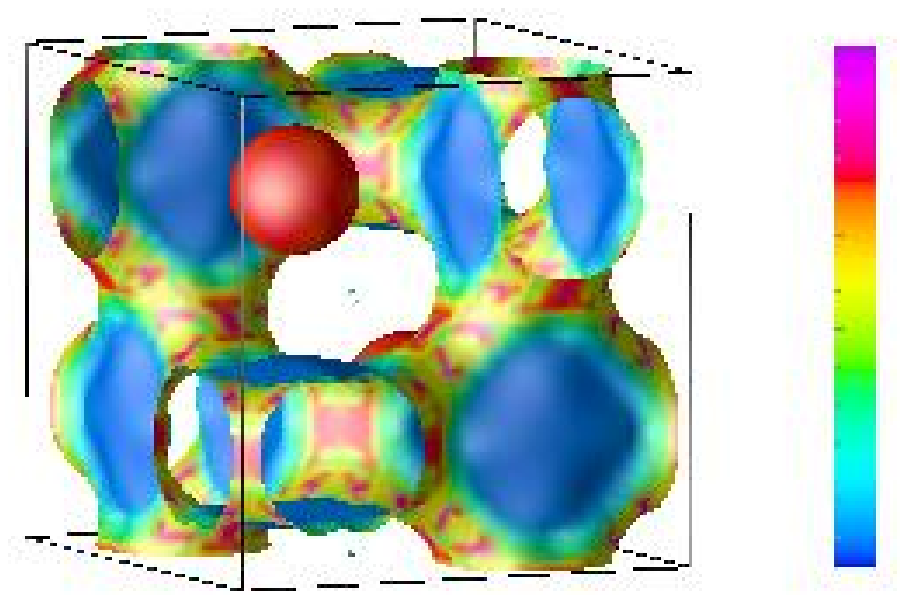,height=0.6\linewidth} }
\vspace{-.2in}
\centerline{
\epsfig{file=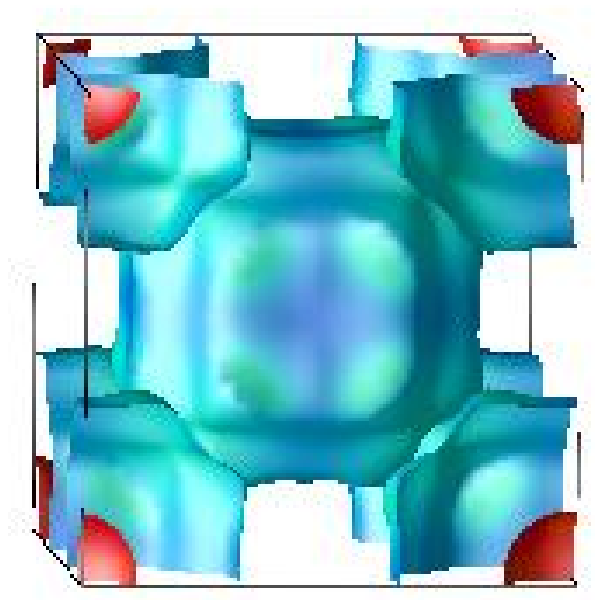,height=0.6\linewidth} }\centerline{
\epsfig{file=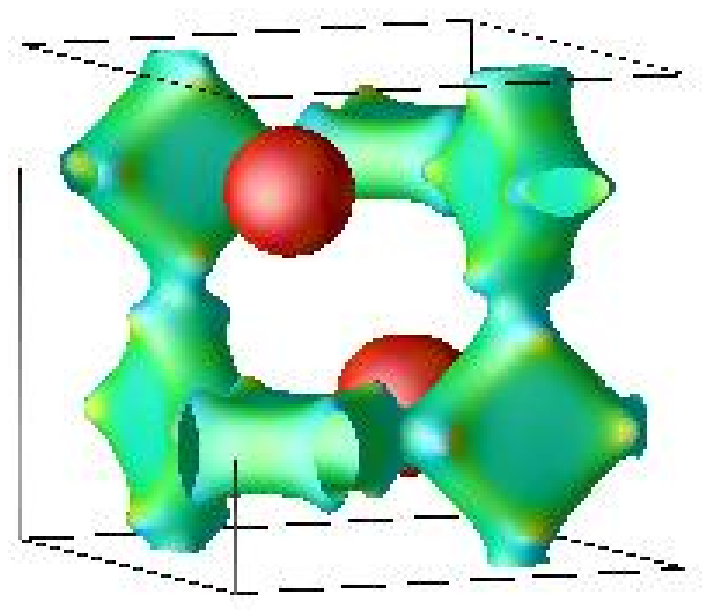,height=0.6\linewidth} }\centerline{
\epsfig{file=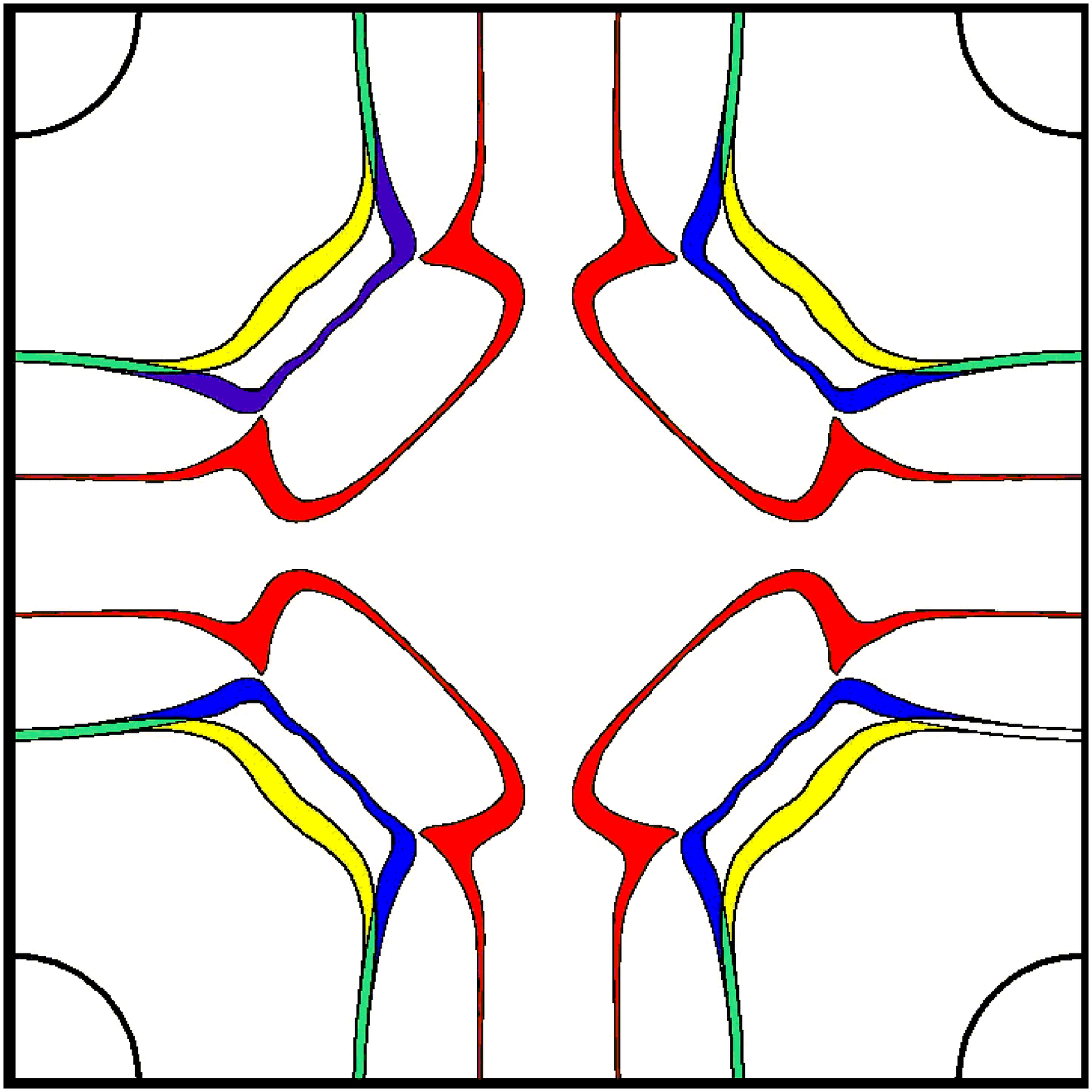,height=0.45\linewidth} }
\setlength{\columnwidth}{3.2in} 
\caption{Fermi surface of ZrZn$_2$. The three upper figures show the nongnatic
Fermi surfaces, colored by the relative percentage of Zn character. The color
bar goes from 5\% to 60\%. The bands shown are, from up down: 
bands 1 and 2 ($\Gamma$-points at the corners), bands 1 and 3 
(L-points at the corners), bands 1 and 4 ($\Gamma$-points at the corners). The bottom panel
shows the Fermi surface 100 ($k_x =0$)
cross-section. The widths of the lines correspond to actual 
exchange splitting for magnetization $M=0.2\ \mu_B$.  } \label{fermi}
\end{figure}

We calculated the band structure of both para- and ferromagnetic ZrZn$%
_{2},$ using the experimental
lattice parameter of 13.858 a.u.\cite{str}.
 The calculations were done in the local spin density
approximation (LSDA) \cite{hl} with the general potential linearized
augmented planewave (LAPW) method. \cite{singh-book}. Well converged
zone samplings and basis sets, including local orbitals \cite%
{singh-lo} to treat the semicore states and relax linearization errors,
were used. 
 The spin-polarized band structure
is shown in Fig.\ref{bands}. The Fermi surface is shown in Fig.%
\ref{fermi}. Our calculations agree well with those published before (Refs.\cite%
{huang,jarlborg} and references therein). The basic electronic
characteristics of the nonmagnetic electronic structure 
are: density of states at the Fermi level, $N(E_{F})=2.43$
states/eV spin f.u., Fermi velocity, $v_{F}=2.5\times 10^{7}$ cm/sec, 
plasma frequency, $\omega _{pl}=4.0$ eV,
Hall concentration $5\times 10^{22}$ e/cm$^{3}$\cite{hall}. The
bands near the Fermi level can be described in the first approximation as
originating from the bonding combination of the $t_{2g}$ Zr orbitals (the
electronic sphere  and the electronic rounded dodecahedron, both around the $%
\Gamma $ point), and the bonding $e_{g}$ orbitals (the
``interconnected pancakes''). The  tubular network
Fermi-surface is a hybridized combination of both types of states. However,
such a view is oversimplified. Comparing the width of the $d$-band of ZrZn$%
_{2},$ and of the pure Zr sublattice (with Zn removed), we observe that the
band width of the latter is about 30\% smaller. Using the well-known
nearest-neighbor tight binding formula, we see that $W_{ZrZn_{2}}\approx 
\sqrt{4t_{Zr-Zr}^{2}+12t_{Zr-Zn}^{2}},$ which gives a rough estimate of  $%
t_{Zr-Zn}\sim 0.5t_{Zr-Zr},$ not surprising, given the smaller radius of Zn.

Although the effect of Zn on the total band width is relatively small (cf.
with the fcc Zr with the same Zr-Zr distance, which has twice larger
bandwidth, $W_{Zr-fcc}\approx \sqrt{12t_{Zr-Zr}^{2}}$\cite{note}), it is not
neglegible, and, interestingly, it is nonuniform over the Brillouin zone.
Specifically, the electron surface around the $\Gamma $ point has more than
50\% Zn character, while the other bands have nearly everywhere less than
20\% of it (Fig.\ref{fermi})\cite{note2}. This leads, in turn, to
nonuniform exchange splitting $\Delta_{xc}({\bf k})$(Fig.\ref{fermi}). For the Fermi
surface near the $\Gamma $ point, the splitting of the Fermi surface is
extremely small, $\delta k_{F}\approx 0.017$ \AA $^{-1}.$ The FFLO
period, $2\pi /\delta k_{F}\approx 360$ \AA , is long enough to allow for
superconductivity in this band with a coherence length $\xi \sim 290$ \AA\ %
(according to Ref.\cite{Peid}). This band has an exceptionally small
$\delta k_F$, but the other bands have small $\delta k_F$ as well. In fact, for
the most of the Fermi surface $\delta k_F\alt 0.07$ \AA $^{-1},$
corresponding to $2\pi /\delta k_{F}\approx 90$ \AA .

The spin-polarized bands, used to produce the lower panel of Fig.\ref{fermi}, are shown in Fig.\ref%
{bands}. These were calculated by fixing the total magnetization to be 0.2 $%
\mu _{B}$/f.u$.$ Fully relaxed calculations converge to the value of 0.7 $%
\mu _{B}$/f.u., nearly 4 times larger than the accepted experimental value
of 0.17-0.20 $\mu _{B}$/f.u. This is somewhat unexpected, for usually
correlation effects, not completely accounted for in LSDA, tend to {\it %
increase} the tendency to magnetism. It  may be that the actual
samples are spatially inhomogeneous, with  magnetic  and
nonmagnetic (and possibly superconducting) regions. Alternatively,
the LSDA may overestimate tendency to magnetism in this
material for some unknown reason.

In vicinity of an itinerant ferromagnetic critical point,
theory predicts a triplet superconductivity, with $T_{c}$ first
growing as one moves in either direction from the critical point, and then
decaying as the spin fluctuations become weaker in either nonmagnetic, or
ferromagnetic phases. Such a behavior has been observed on the ferromagnetic side,
for instance, in UGe$_{2},$ and it was suggested that the difference 
between the longitudinal spin fluctuation spectra in the para- and ferromagnetic phases
makes superconductivity much weaker  and thus
unobservable on the paramagnetic side\cite{K}. So, a triplet state remains a plausible
explanation for the superconductivity in ZrZn$_{2}.$ We want to call
attention, however, to another possibility, namely an FFLO 
inhomogeneous s-wave superconducting state. Both models can successfully
explain the major experimental observations\cite{Peid}: high sensitivity to
sample purity (impurity scattering would make the superconducting
electrons from a band with a small exchange splitting feel a large splitting of the
 other bands),
absence of a specific heat jump (the Fermi surface sections with a large
splitting may have vanishing gaps), disappearence of superconductivity near
the FM critical point (because of the pair-breaking effect of spin
fluctuations). An apparent difference between the two models is that in the latter
superconductivity not only is restored at pressures higher than critical
(this is true for most models of the triplet superconductivity as well), but
also, generically, $T_{c}$ grows much faster on the paramagnetic side than on the
ferromagnetic one. The high pressure data reported in Ref. \cite{Peid} do
not extend far enough beyond the critical point and do not give a definite
answer to whether superconductivity reappears at higher pressures or not.
Unless the strongly interacting
Zr rattling modes would harden substantially, and/or lose their coupling
with electrons, under pressure, the FFLO scenario suggests 
superconductivity
re-entrance at a pressure high enough to suppress the spin fluctuations.
We did check the pressure dependence of the $T_{2g}$ mode, and 
found that under a 
3\% volume compression
it stiffens by 11\%, while it softens by 6\% under a 3\% expansion.

A major question that arises in connection with this alternative is, what
would be the pairing interaction behind the assumed s-wave state? It needs
to be sufficiently strong to make an FFLO state energetically
competitive. To gain more insight in that, we performed lattice dynamics
calculations for the zone center optical modes\cite{phon}. There are 15 such modes at 6
distinctive frequencies: 4 triple-degenerate modes of $T_{1u}$ (two), $T_{2u},$ and $T_{2g}$
modes, one double-degenerate $E_{u}$ mode, and one non-degenerate $A_{1u}$
mode (Table 1). 
The two softest modes  are particularly interesting. Physically,
they correspond to ``rattling'' motion of Zr inside the Zn cage,
hence their softness and the small difference in their frequencies. 
Since these two modes join at the zone boundary, one can see that 
they have little dispersion, as is common for rattling modes.
The even ($T_{2g}$) mode is the only one that can couple with phonons
at the $\Gamma$ point; but as these two modes are essentially local
vibrations, they presumably couple with about the same strength for 
general wave vectors, so the coupling constant computed for a $T_{2g}$
mode should be a reasonable estimate for the coupling constants of all
six rattling  $T_{2g}$ and $T_{1u}$ phonons.  

\vspace{-.2in}
 The coupling constant,
defined as $\lambda _{ep}=2\sum_{{\bf k}\alpha }\delta (\varepsilon _{{\bf k}%
\alpha }-E_{F})(d\epsilon _{{\bf k}\alpha }/du)^{2}/2M\omega ^{2},$ where
$\sum_{{\bf k}\alpha }\delta (\varepsilon _{{\bf k}%
\alpha }-E_{F})$ is the density of states at the Fermi level per spin, 
$M$ is the ionic mass, $\omega$ is the phonon frequency,
and $(d\epsilon _{{\bf k}\alpha }/du)^{2}$, is $\lambda_{ep}=0.115$.
Note that this mode is triple degenerate at $\Gamma ,$ so
the total contribution to $\lambda _{ep}$ from this mode is $\approx $0.35.
However, since we expect that all rattling modes of Zr would couple at a
general point in the Brillouin zone with a similar strength, we estimate the
total contribution from the six rattling modes to be $\approx 0.7$. Finally, the remaining
12 modes (especially the two rather soft $E_u$ modes, 
which we expect to couple less with electons, should bring
a smaller, but finite contribution to the the total $%
\lambda _{ep}$, so we take as a rough estimate $\lambda _{ep}\approx 1.$ 
With a Coulomb pseudopotential $\mu ^{\ast }=0.1,$
the McMillan formula yields $T_{c}\approx 10$ K. This is a reasonable
number since spin fluctuations should reduce $T_{c}$ (and eventually
eliminate superconductivity) in this scenario. This is also consistent with
the transport coupling constant extracted from the linear part of the
resistivity vs. temperature dependence\cite{rho} of ZrZn$_{2}$ single
crystals\cite{res}, $\lambda _{tr}=1.4.$ Assuming that $\lambda _{tr}\approx \lambda
_{ep}+\lambda _{es},$ where $\lambda _{es}$ characterizes coupling with spin
fluctuations, we get $\lambda _{es}\approx 0.4,$ and the modified McMillan
formula then gives $T_{c}=(\omega _{ph}/1.2)\exp [-1.02(1+\lambda
_{ep}+\lambda _{es})/(\lambda _{ep}-\lambda _{es}-\tilde{\mu}^{\ast
})\approx 0.3$ K. Here $\tilde{\mu}^{\ast }=\mu ^{\ast }(1+0.62\lambda
_{ep}+0.62\lambda _{es}).$ On the other hand, the specific heat coefficient
of 47 mJ mol$^{-1}$K$^{-2},$ reported in Ref.\cite{Peid}, corresponds to
mass renormalization 4.1, considerably larger than $(1+\lambda _{ep}+\lambda
_{es})=2.4.$\cite{les}

\begin{figure}[tbp]
\centerline{\epsfig{file=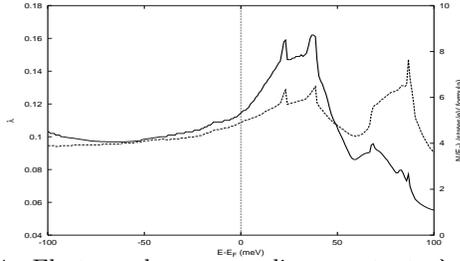,width=0.72\linewidth, height=0.5\linewidth}}
\setlength{\columnwidth}{3.2in} \nopagebreak
\caption{Electron-phonon coupling constant, $\lambda _{ep}$, for one $T_{2g}$ mode
as a function of the Fermi level position. Solid line is $\lambda _{ep}$,
dashed line is density of states.}
\label{lam}
\end{figure}

Another interesting observation is that the topology of the Fermi surface changes 
so sharply with the chemical potential, that the calculated coupling constant 
$\lambda _{ep}$ is very sensitive to the exact position of the Fermi level
in the band structure. In Fig.\ref{lam} we show its dependence on the Fermi
level position. We observe that if the Fermi level were 20 meV,
$\lambda _{ep}$ would be 50\% larger. This unsual sensitivity
is only partially accounted for by the structure of the density of states
(Fig.\ref{lam}). Of course, the tendency to magnetism also
strongly depends on the position of the Fermi level.

\vspace{-.35in}

In conclusion, we report accurate calculations of the electronic
structure and the zone-center phonons in ZrZn$_{2}$ both in para- and
ferromagnetic state. The recently observed superconductivy is compatible with
any of the three scenarios: (1) sample inhomogeneity leading to separation
of superconductivity and magnetism in real space, (2) coexistence of
ferromagnetism and inhomogeneous s-wave superconducting state (Fulde-Ferrell%
-Larkin-Ovchinnikov
state), and (3) triplet spin-fluctuation induced superconductivity. Further
experiments should be able distinguish between these three scenarios, since
they predict rather different thermodynamic behavior: an exponential BCS-like
temperature dependencies vs. a finite residual density of states vs. a power law
due to the gap nodes.

This work was supported by ONR and the ASC supercomputer center.

\begin{table}[tbp]
\caption{Zone center phonon modes in ZrZn$_2$}
\label{tbl}
\begin{tabular}{c|cccccc}
mode & $T_{2g}^a$ & $T_{1u}^b$ & $E_{u}^c$& $T_{1u}^d$ & $A_{1u}^e$ & $T_{2u}^c$ \\
\tableline
character&Zr&mostly Zr&Zn&mostly Zn&Zn&Zn\\
$\omega$ (cm$^{-1}$) & 120 & 133 &142& 182 & 250 &277 \\
\end{tabular}
$^a$ Out-of-phase Zr rattling; $^b$ In-phase Zr rattling; $^c$ Zn breathing
 (the tetrahedra breathe out of phase with each other); $^d $Mixed mode;
$^e$Zn breathing (all three tetrahedra breathe in phase).
\end{table}
\end{multicols}

\begin{references}
\bibitem{note} Note that the hopping in the fcc lattice is $dd\sigma ,$
while in the diamond structure it is an average of $dd\sigma $ and $dd\pi $
hoppings, which gives an additional increase in the bandwidth, and that the
diamond structure posess a pseudogap due to the bonding-antibonding
splitting of the two Zr, thus increasing the density of states at the Fermi
level.
\bibitem{note2} This also can be seen in the shape of this band along the $%
\Gamma $-X line: in the nearest-neighbor Zr-Zr model, this band should be
cosinusoidal, but a noticeable second harmonic component ins clearly seen in
Fig.\ref{bands}. This is due to the Zn-assisted effective second neigbor
hopping.
\bibitem{phon} Standard frozen-phonon technique was employed for
the lattice dynamics calculations; both total energies and forces
were used for constructing the dynamical matrix.
\bibitem{res} Note that the resistivity reported in Ref.\cite{rho} starts to deviate
from the linear behavior already at $T\agt 150$ K. This can be attributed to 
the ``resistivity saturation'' due the proximity to the Ioffe-Regel localization.
Indeed, the mean free path at $T= 150$ K, computed using the LSDA Fermi velocity,
is only twice longer than the lattice parameter.
\bibitem{les}This analysis is based on the assumption that $\lambda _{es}$
is the same above and below the Curie temperature. This is partially justified by the fact that 
$\rho(T)$ does not show any anomaly at $T_C$.\cite{rho}
\end{references}
\end{document}